\input harvmac
\input amssym

\def\unit{\relax{\rm 1\kern-.26em I}}
\def\nada{\relax{\rm 0\kern-.30em l}}
\def\tilde{\widetilde}

%\draftmode

%\def\Omega{\rho,\sigma,\nu  }

\def\CP{{\cal P}}
%% MACROS
\noblackbox
\def\IL{\relax{\rm I\kern-.18em L}}
\def\IH{\relax{\rm I\kern-.18em H}}
\def\IR{\relax{\rm I\kern-.18em R}}
\def\IC{\relax\hbox{$\inbar\kern-.3em{\rm C}$}}
\def\IZ{\relax\ifmmode\mathchoice
{\hbox{\cmss Z\kern-.4em Z}}{\hbox{\cmss Z\kern-.4em Z}}
{\lower.9pt\hbox{\cmsss Z\kern-.4em Z}} {\lower1.2pt\hbox{\cmsss
Z\kern-.4em Z}}\else{\cmss Z\kern-.4em Z}\fi}
\def\CM {{\cal M}}
\def\CN {{\cal N}}

\def\CF {{\cal F}}

\def\CP {{\cal P }}

\def\CO {{\cal O}}

\def\CC {{\cal C}}

%% MORE MACROS

\def\CO {{\cal O}}

\def\CP {{\cal P }}

\def\Tr{{\rm Tr\,}}

\font\manual=manfnt \def\dbend{\lower3.5pt\hbox{\manual\char127}}

\def\IZ{\relax\ifmmode\mathchoice
{\hbox{\cmss Z\kern-.4em Z}}{\hbox{\cmss Z\kern-.4em Z}}
{\lower.9pt\hbox{\cmsss Z\kern-.4em Z}} {\lower1.2pt\hbox{\cmsss
Z\kern-.4em Z}}\else{\cmss Z\kern-.4em Z}\fi}

\def\bar{\overline}

\def\rt2{\sqrt{2}}
\def\irt2{{1\over\sqrt{2}}}

\def\hat{\widehat}
%  \slashchar puts a slash through a character to represent contraction
%  with Dirac matrices. Use \not instead for negation of relations, and use
%  \hbar for hbar.
\def\slashchar#1{\setbox0=\hbox{$#1$}           % set a box for #1
   \dimen0=\wd0                                 % and get its size
   \setbox1=\hbox{/} \dimen1=\wd1               % get size of /
   \ifdim\dimen0>\dimen1                        % #1 is bigger
      \rlap{\hbox to \dimen0{\hfil/\hfil}}      % so center / in box
      #1                                        % and print #1
   \else                                        % / is bigger
      \rlap{\hbox to \dimen1{\hfil$#1$\hfil}}   % so center #1
      /                                         % and print /
   \fi}

\def\foursqr#1#2{{\vcenter{\vbox{
    \hrule height.#2pt
    \hbox{\vrule width.#2pt height#1pt \kern#1pt
    \vrule width.#2pt}
    \hrule height.#2pt
    \hrule height.#2pt
    \hbox{\vrule width.#2pt height#1pt \kern#1pt
    \vrule width.#2pt}
    \hrule height.#2pt
        \hrule height.#2pt
    \hbox{\vrule width.#2pt height#1pt \kern#1pt
    \vrule width.#2pt}
    \hrule height.#2pt
        \hrule height.#2pt
    \hbox{\vrule width.#2pt height#1pt \kern#1pt
    \vrule width.#2pt}
    \hrule height.#2pt}}}}
\def\psqr#1#2{{\vcenter{\vbox{\hrule height.#2pt
    \hbox{\vrule width.#2pt height#1pt \kern#1pt
    \vrule width.#2pt}
    \hrule height.#2pt \hrule height.#2pt
    \hbox{\vrule width.#2pt height#1pt \kern#1pt
    \vrule width.#2pt}
    \hrule height.#2pt}}}}
\def\sqr#1#2{{\vcenter{\vbox{\hrule height.#2pt
    \hbox{\vrule width.#2pt height#1pt \kern#1pt
    \vrule width.#2pt}
    \hrule height.#2pt}}}}

\def\figin{\epsfcheck\figin}\def\figins{\epsfcheck\figins}
\def\epsfcheck{\ifx\epsfbox\UnDeFiNeD
\message{(NO epsf.tex, FIGURES WILL BE IGNORED)}
\gdef\figin##1{\vskip2in}\gdef\figins##1{\hskip.5in}% blank space instead
\else\message{(FIGURES WILL BE INCLUDED)}%
\gdef\figin##1{##1}\gdef\figins##1{##1}\fi}
\def\DefWarn#1{}
\def\figinsert{\goodbreak\midinsert}
\def\ifig#1#2#3{\DefWarn#1\xdef#1{fig.~\the\figno}
\writedef{#1\leftbracket fig.\noexpand~\the\figno}%
\figinsert\figin{\centerline{#3}}\medskip\centerline{\vbox{\baselineskip12pt
\advance\hsize by -1truein\noindent\footnotefont{\bf
Fig.~\the\figno:\ } \it#2}}
\bigskip\endinsert\global\advance\figno by1}

% Load Table package and adjust table size
% file: ruled.tex            TeXsis                  version 2.14
% $Revision: 1.8 $  :  $Date: 91/06/04 14:03:51 $  :  $Author: myers $
%======================================================================*
%  RULED TABLES.  Plain TeX macros for making nice ruled tables.
%
%  The actual guts of the ruled.tex macros are in the file TXSruled.tex,
%  which is loaded at the end of this file.  Here we just include
%  the interline spacing macros from TeXsis since they are also of
%  use in the table making macros.
%
% (C) copyright 1990, 1991 by Eric Myers and Frank E. Paige
%--------------------------------------------------*
% INTERLINE SPACING. (from TeXsis)

                 % synonym for \singlespaced

                 % synonym for \doublespaced

 % synonym for \TrueDoubleSpacing

                 % synonym for \widenspacing
                  % synonym for \widenspacing

% \setRuledStrut creates a vertical strut to hold the interline
%  spacing in ruled tables.  It is defined in TXSruled.tex

%--------------------------------------------------*
% RULED TABLE MACROS:

\input TXSruled.tex

%>>> EOF TXSruled.tex <<<

\thicksize=1pt

\def\Table#1#2{%
   \setbox0=\hbox to 1.5in{\hfil\bf Table~#1:~}%
   \par\hangindent1.5in\hangafter1%
   \noindent\box0 #2%
}%

%\IntriligatorJJ
\lref\IntriligatorJJ{
  K.~A.~Intriligator and B.~Wecht,
  ``The Exact superconformal R symmetry maximizes a,''
Nucl.\ Phys.\ B {\bf 667}, 183 (2003).
[hep-th/0304128].
%%CITATION = hep-th/0304128%%
}

%\LeighEP
\lref\LeighEP{
  R.~G.~Leigh and M.~J.~Strassler,
  ``Exactly marginal operators and duality in four-dimensional N=1 supersymmetric gauge theory,''
Nucl.\ Phys.\ B {\bf 447}, 95 (1995).
[hep-th/9503121].
%%CITATION = hep-th/9503121%%
}

%\HofmanAR
\lref\HofmanAR{
  D.~M.~Hofman, J.~Maldacena,
  ``Conformal collider physics: Energy and charge correlations,''
JHEP {\bf 0805}, 012 (2008).
[arXiv:0803.1467 [hep-th]].
%%CITATION = arXiv:0803.1467%%
}

%\BaggioSNA
\lref\BaggioSNA{
  M.~Baggio, V.~Niarchos and K.~Papadodimas,
  ``Exact correlation functions in SU(2) N=2 superconformal QCD,''
[arXiv:1409.4217 [hep-th]].
%%CITATION = CCTP-2014-18%%
}

%\BaggioIOA
\lref\BaggioIOA{
  M.~Baggio, V.~Niarchos and K.~Papadodimas,
  ``$tt^*$ equations, localization and exact chiral rings in 4d N=2 SCFTs,''
[arXiv:1409.4212 [hep-th]].
%%CITATION = CCQCN-2014-38%%
}

%\ArgyresEH
\lref\ArgyresEH{
  P.~C.~Argyres, M.~R.~Plesser and N.~Seiberg,
  ``The Moduli space of vacua of N=2 SUSY QCD and duality in N=1 SUSY QCD,''
Nucl.\ Phys.\ B {\bf 471}, 159 (1996).
[hep-th/9603042].
%%CITATION = hep-th/9603042%%
}

%\PapadodimasEU
\lref\PapadodimasEU{
  K.~Papadodimas,
  ``Topological Anti-Topological Fusion in Four-Dimensional Superconformal Field Theories,''
JHEP {\bf 1008}, 118 (2010).
[arXiv:0910.4963 [hep-th]].
%%CITATION = arXiv:0910.4963%%
}

%\AsninXX
\lref\AsninXX{
  V.~Asnin,
  ``On metric geometry of conformal moduli spaces of four-dimensional superconformal theories,''
JHEP {\bf 1009}, 012 (2010).
[arXiv:0912.2529 [hep-th]].
%%CITATION = arXiv:0912.2529%%
}

%\DolanZH
\lref\DolanZH{
  F.~A.~Dolan and H.~Osborn,
  ``On short and semi-short representations for four-dimensional superconformal symmetry,''
Annals Phys.\  {\bf 307}, 41 (2003).
[hep-th/0209056].
%%CITATION = hep-th/0209056%%
}

%\GerchkovitzGTA
\lref\GerchkovitzGTA{
  E.~Gerchkovitz, J.~Gomis and Z.~Komargodski,
  ``Sphere Partition Functions and the Zamolodchikov Metric,''
[arXiv:1405.7271 [hep-th]].
%%CITATION = arXiv:1405.7271%%
}

%\KinneyEJ
\lref\KinneyEJ{
  J.~Kinney, J.~M.~Maldacena, S.~Minwalla and S.~Raju,
  ``An Index for 4 dimensional super conformal theories,''
Commun.\ Math.\ Phys.\  {\bf 275}, 209 (2007).
[hep-th/0510251].
%%CITATION = hep-th/0510251%%
}

%\CecottiFI
\lref\CecottiFI{
  S.~Cecotti, A.~Neitzke and C.~Vafa,
  ``R-Twisting and 4d/2d Correspondences,''
[arXiv:1006.3435 [hep-th]].
%%CITATION = arXiv:1006.3435%%
}

%\KomargodskiVJ
\lref\KomargodskiVJ{
  Z.~Komargodski and A.~Schwimmer,
  ``On Renormalization Group Flows in Four Dimensions,''
JHEP {\bf 1112}, 099 (2011).
[arXiv:1107.3987 [hep-th]].
%%CITATION = arXiv:1107.3987%%
}

%\GreenDA
\lref\GreenDA{
  D.~Green, Z.~Komargodski, N.~Seiberg, Y.~Tachikawa and B.~Wecht,
  ``Exactly Marginal Deformations and Global Symmetries,''
JHEP {\bf 1006}, 106 (2010).
[arXiv:1005.3546 [hep-th]].
%%CITATION = arXiv:1005.3546%%
}

%\KomargodskiXV
\lref\KomargodskiXV{
  Z.~Komargodski,
  ``The Constraints of Conformal Symmetry on RG Flows,''
JHEP {\bf 1207}, 069 (2012).
[arXiv:1112.4538 [hep-th]].
%%CITATION = arXiv:1112.4538%%
}

%\ArgyresCN
\lref\ArgyresCN{
  P.~C.~Argyres and N.~Seiberg,
  ``S-duality in N=2 supersymmetric gauge theories,''
JHEP {\bf 0712}, 088 (2007).
[arXiv:0711.0054 [hep-th]].
%%CITATION = arXiv:0711.0054%%
}

%\BeemYN
\lref\BeemYN{
  C.~Beem and A.~Gadde,
  ``The $N=1$ superconformal index for class $S$ fixed points,''
JHEP {\bf 1404}, 036 (2014).
[arXiv:1212.1467 [hep-th]].
%%CITATION = arXiv:1212.1467%%
}

%\XiePUA
\lref\XiePUA{
  D.~Xie and K.~Yonekura,
  ``The moduli space of vacua of N=2 class S theories,''
[arXiv:1404.7521 [hep-th]].
%%CITATION = arXiv:1404.7521%%
}

%\BuicanQLA
\lref\BuicanQLA{
  M.~Buican, T.~Nishinaka and C.~Papageorgakis,
  ``Constraints on Chiral Operators in N=2 SCFTs,''
[arXiv:1407.2835 [hep-th]].
%%CITATION = arXiv:1407.2835%%
}

%\DobrevQV
\lref\DobrevQV{
  V.~K.~Dobrev and V.~B.~Petkova,
  ``All Positive Energy Unitary Irreducible Representations of Extended Conformal Supersymmetry,''
Phys.\ Lett.\ B {\bf 162}, 127 (1985).
}

%\BuicanICA
\lref\BuicanICA{
  M.~Buican,
  ``Minimal Distances Between SCFTs,''
JHEP {\bf 1401}, 155 (2014).
[arXiv:1311.1276 [hep-th]].
%%CITATION = arXiv:1311.1276%%
}

%\GaiottoWE
\lref\GaiottoWE{
  D.~Gaiotto,
  ``N=2 dualities,''
JHEP {\bf 1208}, 034 (2012).
[arXiv:0904.2715 [hep-th]].
%%CITATION = arXiv:0904.2715%%
}

%\XieJC
\lref\XieJC{
  D.~Xie and P.~Zhao,
  ``Central charges and RG flow of strongly-coupled N=2 theory,''
JHEP {\bf 1303}, 006 (2013).
[arXiv:1301.0210].
%%CITATION = DAMTP-2013-1%%
}

%\KomargodskiRB
\lref\KomargodskiRB{
  Z.~Komargodski and N.~Seiberg,
  ``Comments on Supercurrent Multiplets, Supersymmetric Field Theories and Supergravity,''
JHEP {\bf 1007}, 017 (2010).
[arXiv:1002.2228 [hep-th]].
%%CITATION = arXiv:1002.2228%%
}

\lref\ChangSG{
  C.~-M.~Chang and X.~Yin,
  ``Families of Conformal Fixed Points of N=2 Chern-Simons-Matter Theories,''
JHEP {\bf 1005}, 108 (2010).
[arXiv:1002.0568 [hep-th]].
%%CITATION = arXiv:1002.0568%%
}

%\KolUB
\lref\KolUB{
  B.~Kol,
  ``On Conformal Deformations II,''
[arXiv:1005.4408 [hep-th]].
%%CITATION = arXiv:1005.4408%%
}

%\GomisWOA
\lref\GomisWOA{
  J.~Gomis and N.~Ishtiaque,
  ``Kahler Potential and Ambiguities in 4d N=2 SCFTs,''
[arXiv:1409.5325 [hep-th]].
%%CITATION = arXiv:1409.5325%%
}

%\MontonenSN
\lref\MontonenSN{
  C.~Montonen and D.~I.~Olive,
  ``Magnetic Monopoles as Gauge Particles?,''
Phys.\ Lett.\ B {\bf 72}, 117 (1977).
%%CITATION = CERN-TH-2391%%
}

%\AharonyHX
\lref\AharonyHX{
  O.~Aharony, B.~Kol and S.~Yankielowicz,
  ``On exactly marginal deformations of N=4 SYM and type IIB supergravity on AdS(5) x S**5,''
JHEP {\bf 0206}, 039 (2002).
[hep-th/0205090].
%%CITATION = hep-th/0205090%%
}

%\KolZT
\lref\KolZT{
  B.~Kol,
  ``On conformal deformations,''
JHEP {\bf 0209}, 046 (2002).
[hep-th/0205141].
%%CITATION = hep-th/0205141%%
}

%\TachikawaTQ
\lref\TachikawaTQ{
  Y.~Tachikawa,
  ``Five-dimensional supergravity dual of a-maximization,''
Nucl.\ Phys.\ B {\bf 733}, 188 (2006).
[hep-th/0507057].
%%CITATION = hep-th/0507057%%
}

%\BarnesBW
\lref\BarnesBW{
  E.~Barnes, E.~Gorbatov, K.~A.~Intriligator and J.~Wright,
  ``Current correlators and AdS/CFT geometry,''
Nucl.\ Phys.\ B {\bf 732}, 89 (2006).
[hep-th/0507146].
%%CITATION = hep-th/0507146%%
}

\Title{\vbox{\baselineskip12pt\hbox{RU-NHETC-2014-16}
		}}{Compact Conformal Manifolds}

%\smallskip
\centerline{Matthew Buican\footnote{$^{\dagger}$}
{buican@physics.rutgers.edu} and Takahiro Nishinaka\footnote{$^{*}$}{nishinaka@physics.rutgers.edu}}
\smallskip
\bigskip
\centerline{{\it NHETC and Department of Physics and
Astronomy}}\vskip -.04in
\centerline{{\it Rutgers University, Piscataway, NJ 08854, USA}}
%\smallskip

\vskip .95cm \noindent In this note we begin a systematic study of compact conformal manifolds of SCFTs in four dimensions (our notion of compactness is with respect to the topology induced by the Zamolodchikov metric). Supersymmetry guarantees that such manifolds are K\"ahler, and so the simplest possible non-trivial compact conformal manifold in this set of geometries is a complex one-dimensional projective space. We show that such a manifold is indeed realized and give a general prescription for constructing complex $N$-dimensional projective space conformal manifolds as certain small $\CN=2\to\CN=1$ breaking deformations of strongly interacting $\CN=2$ SCFTs. In many cases, our prescription reduces the construction of such spaces to a study of the $\CN=2$ chiral ring. We also give an algorithm for constructing more general compact spaces of SCFTs.

\Date{October 2014}

\newsec{Introduction}
A conformal manifold, $\CM$, is a family of Conformal Field Theories (CFTs) that are connected to each other by exactly marginal deformations. More precisely, given some CFT, $\CP$, there may be exactly marginal deformations
\eqn\exactmarginal{
\delta S=\int d^dx\lambda^i\CO_i~,
}
that take us from $\CP$ to some nearby CFT, $\CP'$. Furthermore, $\CM$ is endowed with a positive-definite metric
\eqn\Zamometric{
g_{ij}(\CP)\equiv x^{2d}\cdot\langle\CO_i(x)\CO_j(0)\rangle_{\CP}~,
}
called the Zamolodchikov metric.

Conformal manifolds are more controlled laboratories in which to study phenomena like emergent symmetry \GreenDA\ and dualities (e.g., \refs{\MontonenSN,\GaiottoWE,\ArgyresCN}) that also occur in theories with broken conformal symmetry (e.g., real-world QCD). Understanding general properties of these spaces may therefore be worthwhile.

In this note, we begin a study of compact conformal manifolds in four space-time dimensions (note that in three dimensions, weakly coupled compact conformal manifolds have been constructed in \ChangSG). A necessary and sufficient condition for a finite-dimensional $\CM$ to be compact is that it is complete (i.e., every Cauchy sequence of points in $\CM$ converges in $\CM$) and that the manifold has finite diameter (i.e., the supremum of geodesic distances measured using the Zamolodchikov metric is finite).

Most known conformal manifolds in four dimensions are either non-compact or consist of a single point. A particularly well-known example of the former is the space of exactly marginal gauge coupling(s) of $\CN=4$ Super Yang-Mills (SYM). Specializing to the case of a single gauge group, these manifolds have an exact Zamolodchikov metric
\eqn\ZamoNeqfour{
g_{\tau\bar\tau}d\tau d\bar\tau\sim{1\over({\rm Im}\tau)^2}d\tau d\bar\tau~, \ \ \ \tau={\theta\over2\pi}+i{4\pi\over g^2}~.
}
The zero gauge coupling limit, $\tau=i\infty$ ($g=0$), is infinitely far away, and so, by our above discussion, the conformal manifold of an $\CN=4$ SYM theory has a logarithmically divergent diameter. In fact, similar statements hold for many known theories with extended SUSY, since one often finds cusps where free gauge fields emerge and the Zamolodchikov metric takes the form in \ZamoNeqfour\ modulo small corrections (see \refs{\BaggioIOA, \BaggioSNA} for an interesting computation of these corrections).\foot{We can often compactify these conformal manifolds. However, the resulting metric is not the Zamolodchikov metric (also, unlike in our examples below, the number of stress tensors changes discontinuously along the conformal manifold).} At the opposite extreme, we can consider an isolated CFT. A classic example of such a theory is QCD in the Banks-Zaks phase. In this case, there are no exactly marginal deformations, and the conformal manifold is trivial: it is just a point.

Since our goal is to begin a systematic and controlled study of compact conformal manifolds, we will focus on superconformal field theories (SCFTs) in this note. By the results of \AsninXX, such manifolds are necessarily K\"ahler. Therefore, the simplest (lowest dimension and simplest topology) non-trivial compact supersymmetric conformal manifold that we can imagine constructing is a ${\bf C}P^1$.

The main results of this note are {\bf(i)} to give an algorithm for constructing compact conformal manifolds in conformal perturbation theory in four dimensions and {\bf(ii)} to show that ${\bf C}P^1$ conformal manifolds are indeed realized: they can be naturally constructed in conformal perturbation theory via certain explicit $\CN=2\to\CN=1$ breaking deformations of any of the infinite set of $(A_{2^n-1}, D_{2p})$ Argyres-Douglas (AD) theories \CecottiFI\ with integers $n\ge2$ and $p\gg2^n$.

Moreover, our arguments often reduce the search for ${\bf C}P^{N-1}$ conformal manifolds to a problem in studying the $\CN=2$ chiral ring (by which we mean the ring of operators annihilated by all of the anti-chiral supercharges of $\CN=2$ SUSY). We claim that a ${\bf C}P^{N-1}$ conformal manifold can be realized in conformal perturbation theory by a set of $\CN=2\to\CN=1$ breaking deformations of an infinite sub-family of $\CN=2$ SCFTs, $\left\{\CT_{\rho}\right\}$, labeled by a positive integer, $\rho$, with $\rho$ sufficiently large and each corresponding $\CT_{\rho}$ satisfying:

\smallskip
{\item {\bf(a)}} $\CT_{\rho}$ has no $\CN=2$ chiral primaries of dimension two (i.e., it is isolated as an $\CN=2$ SCFT).

{\item {\bf(b)}} There is a positive real number, $\kappa$, such that  $\CT_{\rho}$ has $N$ $\CN=2$ Lorentz scalar chiral primaries of dimension $2-\epsilon$, with $\epsilon\sim\CO(\rho^{-\kappa})\ll1$.

{\item {\bf(c)}} There is a positive real number, $\kappa'$, such that the conformal $c$ anomaly of $\CT_{\rho}$ is bounded as $c\gtrsim\CO(\rho^{\kappa'})$. Alternatively, we can rephrase this requirement using the $a$-theorem \refs{\KomargodskiVJ,\KomargodskiXV} and the Hofman-Maldacena bounds \HofmanAR\ and demand that, if the theory has a Coulomb branch, $\CM_{\CC}$, then its dimension scales as ${\rm dim}(\CM_{\CC})\sim\CO(\rho^{\kappa'})\gg1$.

\smallskip
\noindent
We will see below that in the case of the $(A_{2^n-1}, D_{2p})$ theories, $\rho=p$, while $\kappa=\kappa'=1$.

Since the conformal manifolds we will construct are visible in conformal perturbation theory (i.e., the diameters are parametrically small), we will have control over their global properties. As we will see, although these conformal manifolds only have $\CN=1$ SUSY, they behave in certain respects like $\CN=2$ conformal manifolds. Whether we can actually realize compact conformal manifolds in extended SUSY itself is an interesting problem that we leave for the future.

The plan of the paper is as follows. We first present our general algorithm for constructing compact conformal manifolds in conformal perturbation theory. Next, we study the $(A_3, D_{2p})$ theory and find a set of $\CN=2\to\CN=1$ breaking deformations that give rise to a small ${\bf C}P^1$ conformal manifold. We then comment on the more general case of $(A_{2^n-1}, D_{2p})$. Finally, we conclude with some discussion of open problems.

\newsec{Compact Conformal Manifolds in Four Dimensions}
In order to understand how compact conformal manifolds can naturally appear in conformal perturbation theory, it is useful to briefly review the criterion for the existence of exactly marginal deformations proven in \GreenDA.\foot{This criterion was originally suggested for general $\CN=1$ four-dimensional SCFTs in \KolZT\ (see also \KolUB).} Let us start from some reference SCFT, $\CP$. Unitarity guarantees that a marginal deformation must be a shift in the superpotential. Let us take a small deformation of the form
\eqn\defW{
\delta W=\lambda^i\CO_i~,
}
where the $\CO_i$ are chiral primaries of dimension three.

In general, not all deformations of the type \defW\ are exactly marginal. In order to distinguish the exactly marginal deformations, let us study the chiral-anti-chiral OPE
\eqn\OOOPE{
\CO_i(x)\CO^{\dagger}_{\bar j}(0)={\delta_{i\bar j}\over |x|^6}+{1\over |x|^4}T_{i\bar j}^AJ_A(0)+\cdots~,
}
where the $J_A$ are the moment maps of the flavor symmetries (with corresponding charges, $Q_A$) at $\CP$, and the ellipses include descendants and primaries of higher dimension that are not directly relevant to the discussion at hand. The coefficients multiplying the moment maps measure the flavor charges of the marginal chiral primaries
\eqn\TAdef{
T^A_{i\bar j}=4\pi^2\tau^{AB}(t_B)_i^k\delta_{k\bar j}~, \ \ \tau_{AB}=(2\pi)^4\cdot |x|^4\cdot\langle J_A(x)J_B(0)\rangle>0~,\ \ [Q_A,\CO_i]=-(t_A)_i^j\CO_j~.
}

Since there are no singular terms in the chiral-chiral OPE, we can work in a holomorphic renormalization scheme with the superpotential unrenormalized and the K\"ahler potential deformed by
\eqn\Krenorm{
\delta K=Z^A(\lambda, \bar\lambda, \mu)\cdot J_A~,
}
where $\mu$ is the short-distance cutoff. Now, let us define
\eqn\Za{
D^A\equiv{\partial\over\partial\log\mu}Z^A=2\pi^2\lambda^iT^A_{i\bar j}\bar\lambda^{\bar j}+\cdots=8\pi^4\tau^{AB}\lambda^i(t_B)^k_i\bar\lambda_k+\cdots~,
}
where we have used the Zamolodchikov metric, $\delta_{k\bar m}$, to lower the index of $\bar\lambda^{\bar m}$. Clearly, the exactly marginal deformations are precisely those that satisfy
\eqn\exactmarg{
D^A=0~.
} 
Note that the beta functions for the physical couplings, $\lambda_{\rm phys}^i=\lambda^i+{1\over2}Z^A(t_A)_j^i\lambda^j$, are
\eqn\betafn{
\beta^i={1\over2}\lambda^j(t_A)_j^i{\partial\over\partial\log\mu}Z^A=4\pi^4\lambda^j(t_A)_j^i\tau^{AB}\lambda^k(t_B)_k^{\ell}\bar\lambda_{\ell}+\cdots~.
}
As a result, small deformations that are not exactly marginal are marginally irrelevant.

As emphasized in \GreenDA, the form of \Za\ is suggestive of a \lq\lq $D$-term" for the global symmetries, and \exactmarg\ is suggestive of a $D$-flatness condition (we can arrive at the condition \exactmarg\ by extremizing the potential $V=2\pi^4\cdot D^AD_A\ge0$). Indeed, if we think of the couplings as background fields transforming under global symmetries coupled to background gauge fields, then we can think of \Za\ as a $D$-term for these symmetries (with $V$ a $D$-term potential). In particular, we see that, at least locally around $\CP$, we can describe the conformal manifold as a symplectic quotient $\CM=\left\{\lambda^i\right\}/G^{\bf C}$ (see also the discussion in \KolUB), where $G^{\bf C}$ is the complexified global symmetry group (we use the fact that acting on the $D$-flat $\lambda^i$ by the symmetry group $G$ does not move us along $\CM$ since the difference between the theory before and after the symmetry rotation is an irrelevant operator \GreenDA). In theories with supergravity duals, the conformal manifold gets mapped to the moduli space of vacua of the gravity theory, and the above construction has a natural dual interpretation in terms of an action on the massless fields (see, for example, the relevant sections in \refs{\KolZT,\AharonyHX\TachikawaTQ-\BarnesBW}). Note, however, that the discussion above (which follows \GreenDA) is more general and applies to any SCFT (with or without a dual).

Therefore, a chiral operator fails to be marginal if it pairs up with a flavor symmetry current to form a long multiplet.\foot{This result can also be explained directly in terms of the recombination rules of superconformal representation theory \BeemYN.} Note that if the theory has $\CN=2$ SUSY, then it must be the case that all $\CN=2$-preserving marginal deformations are exactly marginal (this statement essentially follows from the above discussion and the OPE analysis in \BuicanICA, which shows that flavor symmetry moment maps cannot appear in the chiral-anti-chiral OPE of the marginal primaries), but this is not generically true in $\CN=1$ theories (a simple example of this latter statement is $SU(N_c)$ SQCD with $N_f=2N_c$ \refs{\LeighEP, \GreenDA}).\foot{Actually, we will see that this statement remains true along the $\CN=1$ compact conformal manifolds that we get from parametrically small deformations of the $\CN=2$ SCFTs we consider below.}

Typically, it is not straightforward to use the $D$-flatness condition \exactmarg\ to determine whether a particular conformal manifold is compact or not, since we are performing a local analysis around $\CP$. One simple situation in which we can see a compact conformal manifold in conformal perturbation theory is if we imagine that our D-term equations are deformed by a parametrically small background \lq\lq FI" term, $\xi$ (such terms have a natural interpretation in a symplectic quotient). More precisely, let us suppose our theory has a single $U(1)$ flavor symmetry and that the corresponding D-term is deformed as
\eqn\FIdef{
D\to D-\xi~, \ \ \ 0<\xi\ll1~,
}
where $\xi$ does not depend on the couplings (without loss of generality, we take $\xi>0$). If the couplings all have charges of the same sign (with a sign resulting in a positive coupling-dependent contribution in \FIdef) and these charges are parametrically larger than $\xi$ in absolute value, then it is clear that \FIdef\ describes a compact space in conformal perturbation theory around $\CP$.

A simple way to engineer a conformal manifold of this type is to start from an isolated SCFT\foot{As we will see below, this condition is not strictly necessary.} with various chiral primaries, $\CO_i$, of dimension $3-\epsilon_i$  (where $\epsilon_i\ll1$) and a $G=U(1)$ global symmetry group. We can deform the superpotential by
\eqn\Wdef{
\delta W=\lambda^i\mu^{\epsilon_i}\CO_i~,
}
where the $\lambda^i$ are dimensionless. The beta functions now read
\eqn\betafnui{
\beta^i=-\epsilon_i\lambda^i+4\pi^4\lambda^i\alpha_i\tau_G^{-1}\sum_j\lambda^j\alpha_j\bar\lambda_j+\cdots~,
}
where the first terms are the tree-level contributions to the beta functions, the $\alpha_i$ are the charges of the $\lambda_i$ under $G$, and $\tau_G$ is the two-point function of the $G$ symmetry current. If the above beta functions have a solution with parametrically small $\lambda^i=\lambda^i_*\ne0$ in the IR, then it should be the case that the superpotential \Wdef\ preserves an $R$ symmetry along the RG flow.\foot{More precisely, as long as there are no approximately conserved flavor symmetries with currents, $j^A_{\mu}$, of dimension $3+\epsilon_A'$ for $0<\epsilon_A'\lesssim\epsilon_i$, we expect that this $R$-symmetry exists and that it flows to the IR superconformal one. In the $\CN=2\to\CN=1$ breaking examples below, even if there are such $j_{\mu}^A$, they cannot mix with the IR superconformal $R$-current (this result follows from the discussion in the appendix of \BuicanICA).} In particular, it follows that $\alpha_i=\alpha\cdot\epsilon_i$, for some universal $\alpha\ne0$. We can rescale the symmetry current multiplet $J_G=\alpha \cdot U$ (so that the $U$ charge of $\CO_i$ is $\epsilon_i$, where $U$ is the operator originally introduced in \KomargodskiRB) and find
\eqn\betafnuii{
D_U=-\xi+D={\tau_U\over 4\pi^4}(\epsilon_i\lambda^i)^{-1}\beta^i=-{\tau_U\over4\pi^4}+\sum_j\lambda^j\epsilon_j\bar\lambda_j+\cdots~, \ \ \ \tau_U={\tau_G\over\alpha^2}~.
}
Note that we can define the potential $V=2\pi^4D^UD_U$, obtain the beta function as the gradient of $V$, and find the IR theory by studying the $D$-flat directions in the presence of the background FI term. Assuming that there are no other exactly marginal deformations at long distances, the IR conformal manifold is compact with complex dimension $N-1$, where $N$ is the number of independent deformations in \Wdef, and the IR couplings are characterized by
\eqn\satisfy{
\Big\{\lambda^i\Big|\sum_j\lambda^j_*\epsilon_j\bar\lambda_{*j}={\tau_U\over4\pi^4}\Big\}\Big/U(1)~,
}
where the $U(1)$ action is generated by $U$. The solutions \satisfy\ correspond to directions in the kernel of the Hessian, $\partial_i\partial_{\bar j}V$. If the $\epsilon_j$ are all equal, then we get a ${\bf C}P^{N-1}$. Otherwise, we find a weighted projective space.

As long as $\tau_U\ll\epsilon_i$, we expect that the IR is indeed described by a compact conformal manifold, $\CM_c$, and that higher-order corrections (as well as corrections from higher-dimensional operators) will not render $\CM_c$ non-compact (i.e., they won't change the topology of the conformal manifold) or push the theory non-perturbatively far from $\CP$. We will perform a consistency check of this picture in our examples below when we compute the exact IR superconformal $R$-current. Note that in some cases, the IR solutions to the vanishing of the beta functions \betafnui\ may only exist for a subset of the $\lambda_i\ne0$.

In a more general setting, starting from some UV SCFT, $\CP$, with a global symmetry group $G$ (not necessarily $U(1)$), the beta functions are given by
\eqn\betafn{
\beta^i=-\epsilon_i\lambda^i+4\pi^4\lambda^j(t_A)_j^i\tau^{AB}\lambda^k(t_B)_k^{\ell}\delta_{\ell \bar m}\bar\lambda^{\bar m}+\cdots~.
}
In this case, we can think of the beta functions as gradients of the potential $V=\kappa-\sum_i\epsilon_i\lambda^i\bar\lambda_i+2\pi^4(\lambda^j(t_A)_j^i\bar\lambda_i)\tau^{AB}(\lambda^k(t_B)_k^{\ell}\bar\lambda_{\ell})+\cdots$, where $\kappa$ is a constant.

In the examples below, we will discuss UV theories with a global symmetry group $G\supset U(1)$ and many relevant chiral primaries of dimension approximately three.  Instead of finding the full set of solutions to \betafn\ in these theories, we will turn on a subset of the relevant deformations that preserve a common $R$-symmetry and break a particular $U(1)\subset G$ symmetry. The RG evolution of these couplings is governed by the simpler beta functions in \betafnuii, and the compact conformal manifolds we find in the IR are described by \satisfy\ for an appropriately defined $U$. In order to verify that our conformal manifolds are indeed compact in the IR and of the form we claim, we will show that the various other IR chiral primaries that have UV dimension approximately three do not become exactly marginal in the IR (note that any new IR primaries of dimension three that might emerge in the examples below would necessarily be marginally irrelevant since they would be neutral under a preserved $U(1)$ flavor symmetry we will describe in detail and would be charged under some emergent symmetry). While it would be interesting to study the full set of solutions to \betafn\ in these examples, such a discussion is beyond the scope of this note.

\subsec{The Role of $\CN=2$ SUSY}
As we will explain, $\CN=2$ SCFTs naturally have---from the point of view of an $\CN=1\subset\CN=2$ sub-algebra---sectors of operators that are charged only under $U(1)$ flavor symmetries. These theories are then candidates for implementing the mechanism described above. Indeed, recall that SCFTs with $\CN=2$ SUSY have a $U(1)_R\times SU(2)_R\times \CF$ bosonic internal symmetry, where $U(1)_R\times SU(2)_R$ is the superconformal $R$-symmetry, and $\CF$ is the $\CN=2$ flavor symmetry (i.e., it commutes with the $\CN=2$ superconformal algebra, and, by our conventions, the corresponding currents do not sit in multiplets with higher-spin symmetry currents). From the $\CN=1$ point of view, the linear combination
\eqn\Uijflavor{
J=R_{\CN=2}-2I_3~,
}
is a universal $U(1)$ flavor symmetry of the theory (it commutes with the $\CN=1\subset\CN=2$ superconformal algebra; $I_3$ is the Cartan generator of $SU(2)_R$, and $R_{\CN=2}$ is the $U(1)_R$). Such theories often have a sector of primaries that are charged under $J$ and are singlets under $\CF$: the $\CN=2$ chiral primaries, $\CO_i$ (these operators are annihilated by all the anti-chiral supercharges). Indeed, unitarity implies
\eqn\JchargeO{
J(\CO_i)=2D(\CO_i)\ge2~,
}
where $D(\CO_i)$ is the scaling dimension of $\CO_i$ (the fact that the $\CO_i$ are $\CF$-neutral can be derived by studying the $\CO_i^{\dagger}\CO_i$ OPE \BuicanQLA). Furthermore, we can often compute the scaling dimensions of the $\CO_i$ exactly from a Seiberg-Witten or Calabi-Yau description of the theory.

In the next subsection, we will study a set of relevant $\CN=2\to\CN=1$ breaking deformations that couple a free chiral multiplet to the $(A_3, D_{2p})$ SCFT \CecottiFI\ via a composite operator built out of bilinears of the free chiral scalar and the $\CN=2$ chiral operators discussed above. We will show how a small ${\bf C}P^1$ conformal manifold appears in the IR. We then consider generalizing to cases in which we replace the $(A_3, D_{2p})$ theory with an $(A_{2^n-1}, D_{2p})$ theory ($n>2$).

\subsec{A ${\bf C}P^1$ Conformal Manifold}
Consider the $(A_3, D_{2p})$ Argyres-Douglas SCFT (for $p\gg1$) \CecottiFI\ along with a decoupled chiral multiplet, $\Phi$,  (although it is not important, we will think of this chiral multiplet as being part of a free $\CN=2$ $U(1)$ multiplet). While the $(A_3, D_{2p})$ theory apparently does not admit a Seiberg-Witten description, and many of its properties remain mysterious, we can derive its $\CN=2$ chiral spectrum (and the fact that it has a rank two $\CN=2$ flavor symmetry, $\CF$) from the  Calabi-Yau equation $x^4+y^2+s^{2p-1}+st^2=0$.\foot{Deforming the theory by an $\CN=2$ chiral operator corresponds to adding a lower-order monomial to the polynomial $x^4 + y^2 + s^{2p-1} + st^2$. The spectrum of $\CN=2$ chiral operators is therefore derived by studying all possible deformations of the polynomial. The scaling dimensions of the deformations are determined so that the holomorphic 3-form on the Calabi-Yau threefold has dimension one \CecottiFI .} It turns out that the $\CN=2$ chiral operators have scaling dimensions
\eqn\scalingdims{\eqalign{
&D(\CO_{0,\ell})={4(2p-\ell-1)\over2p+1}~, \ \ \ 0\le\ell\le{\rm floor}\left({6p-5\over4}\right)~, \cr &D(\CO_{1,\ell})={6p-4\ell-3\over 2p+1}~, \ \ \ 0\le\ell\le p-2~,\cr &D(\CO_{2,\ell})={4p-4\ell-2\over2p+1}~, \ \ \ 0\le\ell\le{\rm floor}\left({2p-3\over4}\right)~,\cr &D(\CO_{0})={4p\over2p+1}~. 
}}
As explained above, the operators in \scalingdims\ are neutral under $\CF$ but are charged under the $U(1)$ $J$ symmetry defined in \Uijflavor\ with charges given by twice their scaling dimensions (see \JchargeO). We refer to the particular $J$ symmetry acting on the $(A_3, D_{2p})$ sector and leaving the $\Phi$ multiplet invariant as $J_{(A_3, D_{2p})}$. We refer to the $J$ symmetry acting on $\Phi$ but leaving the $(A_3, D_{2p})$ sector invariant as $J_{\Phi}$. Note that \scalingdims\ implies that the $(A_3, D_{2p})$ theory satisfies properties {\bf(a)}, {\bf(b)}, and {\bf(c)} discussed in the introduction with $N=2$, $\rho=p$, and $\kappa=\kappa'=1$ (we will explain below precisely why $c\gtrsim\CO(p)$).

Let us now turn on the following almost-marginal relevant deformation
\eqn\deformation{
\delta W=\lambda^1\mu^{\epsilon}\cdot\Phi\CO_{0,p-1}+\lambda^2\mu^{\epsilon}\cdot\Phi\CO_0~,
}
where $D(\Phi\CO_{(0,p-1)})=D(\Phi\CO_0)=3-\epsilon$, with $\epsilon={2\over1+2p}\sim{1\over p}$. This deformation breaks $\CN=2\to\CN=1$ and breaks the $\CN=1$ bosonic flavor and $R$-symmetry as follows 
\eqn\flavorsymmbr{
U(1)_{\tilde R}\times U(1)_{J_{\Phi}}\times U(1)_{J_{(A_3, D_{2p})}}\times \CF\to U(1)_R\times U(1)_F\times \CF~,
}
where $U(1)_{\tilde R}$ is the UV $\CN=1\subset\CN=2$ superconformal $R$-symmetry, and $U(1)_R$ is a remaining $R$-symmetry. Turning on any other relevant deformations necessarily breaks \flavorsymmbr\ further. Note that the unbroken $U(1)_F$ symmetry is the linear combination $U(1)_F=-{4p\over2p+1}U(1)_{J_{\Phi}}+U(1)_{J_{(A_3, D_{2p})}}$. Defining the moment map $U={1\over256p^2+9c(1+2p)^2}\left(9c(1+2p)\cdot J_{\Phi}+64p\cdot J_{(A_3, D_{2p})}\right)$ (where $J_{\Phi}$ and $J_{(A_3, D_{2p})}$ are the moment maps for the symmetries of the same name, and $c$ is the central charge of the $(A_3, D_{2p})$ theory), we find that the IR theory is described by
\eqn\AthreeCM{
\Big\{\lambda^i\Big||\lambda^1_*|^2+|\lambda^2_*|^2={1+2p\over2}\cdot{\tau_U\over4\pi^4}={9c(1+2p)\over2\pi^4(256p^2+9c(1+2p)^2)}\lesssim{1\over4\pi^4p}\ll1\Big\}\Big/U(1)~,
}
where we have used the fact that $\tau_J={9\over4}c$, and the $U(1)$ action is generated by $U$.\foot{For an $\CN=1$ theory, we take $c=3\Tr\tilde R^3-{5\over3}\Tr\tilde R$, where $\tilde R={1\over3}R_{\CN=2}+{4\over3}I_3$ is the $\CN=1$ superconformal $R$-symmetry.} Therefore, we see that the IR conformal manifold is $\CM_c= {\bf C}P^1$, and its diameter, ${\rm diam}(\CM_c)$, is at most $\CO(p^{-{1\over2}})$. Note that even though the IR conformal manifold has $\CN=1$ SUSY, it behaves in at least one interesting way as an $\CN=2$ conformal manifold: the flavor symmetry, $U(1)_F\times \CF$, is an invariant of the manifold.\foot{By flavor symmetry, we again note that we mean a symmetry that commutes with the superconformal algebra and does not sit in a multiplet with higher-spin symmetries. Note also that $\CN=2$ flavor symmetries can only emerge when a new sector appears. However, these emergent symmetries are somewhat special, since they are arbitrarily weakly gauged.} Furthermore, it is crucial we satisfy property {\bf(b)} in the introduction with $N=2$ in order to find a ${\bf C}P^1$ (one of the corresponding two chiral operators gets eaten by the $U$ mutliplet and becomes a normal direction).

In order to establish that we in fact have a small ${\bf C}P^1$ conformal manifold in the IR, we should check that there are no additional exactly marginal deformations at long distances. Therefore, we should check that all the UV $\CN=1$ chiral primaries with dimension close to three do not become exactly marginal in the IR. As we proceed to demonstrate this fact below, we will highlight where conditions {\bf(a)} and {\bf(c)} mentioned in the introduction enter into our analysis.

To begin, first note that the $\CN=1$ primaries of interest may either be $\CN=2$ primaries or $\CN=2$ descendants. In either case, the $\CN=1$ primaries are embedded in short multiplets of $\CN=2$ SUSY. The corresponding $\CN=2$ primaries have dimensions determined by their $R$-symmetry quantum numbers as follows \refs{\DobrevQV,\DolanZH}
\eqn\scalingdimsii{
D={1\over2}R_{\CN=2}+2j_R~, \ \ \ R_{\CN=2}=0\ {\rm or} \ R_{\CN=2}\ge2(1+j_1)~,
}
where $j_R$ is the $SU(2)_R$ spin, and $j_1$ is the left-handed spin.

From the above discussion, it is clear that each UV chiral operator of dimension close to three must fall into one of the following categories (the quantum numbers are with respect to the UV superconformal $R$-symmetry, and the dimensions are determined by \scalingdimsii)

\item{\bf(i)} An $\CN=2$ Lorentz scalar chiral primary operator with $j_R=0$ and $R_{\CN=2}\sim6$.

\item{\bf(ii)} An $\CN=2$ Lorentz scalar primary with $j_R={1\over2}$ and $R_{\CN=2}\sim4$.

\item{\bf(iii)} An $\CN=2$ Lorentz scalar primary with $j_R=1$ and $R_{\CN=2}\gtrsim2$.

\item{\bf(iv)} An $\CN=2$ Lorentz scalar primary with $j_R={3\over2}$ and $R_{\CN=2}=0$.

\item{\bf(v)} An $\CN=2$ descendant, $(Q^2)^2\CO$, with $\CO$ an $\CN=2$ Lorentz scalar chiral primary with $j_R=0$ and $R_{\CN=2}\sim4$.

\item{\bf(vi)} An $\CN=2$ descendant, $Q^{2\alpha}\CO_{\alpha}$, with $\CO_{\alpha}$ a chiral primary of spin $\left({1\over2},0\right)$, $j_R=0$, and $R_{\CN=2}\sim5$.

\item{\bf(vii)} An $\CN=2$ descendant, $Q^{2\alpha}\CO_{\alpha}$, with $\CO_{\alpha}$ a chiral primary of spin $\left({1\over2},0\right)$, $j_R={1\over2}$, and $R_{\CN=2}\gtrsim3$.\foot{Such operators coming from primaries of $R_{\CN=2}<3$ are forbidden by unitarity. Furthermore, in the case $R_{\CN=2}=3$, we have $Q^{2\alpha}\CO_{\alpha}=0$. Indeed, since $\CO_{\alpha}$ saturates the unitarity bound $R_{\CN=2}(\CO)=3=2(1+j_1(\CO))$, it is annihilated by the Lorentz spin zero contraction with $Q^{2}_{\beta}$. An example of such an operator is $\CO_{\alpha}=\lambda_{\alpha}\tilde\phi-\tilde\lambda_{\alpha}\phi$, where $\phi$, $\tilde\phi$ are primaries for two different free $U(1)$ vector multiplets, and $\lambda_{\alpha}\sim Q_{\alpha}^2\phi|$, $\tilde\lambda_{\alpha}\sim Q_{\alpha}^2\tilde\phi|$ are the corresponding gauginos.}

Given this list of operators, we can compute the corresponding set of IR superconformal $R$-charges, $\tilde R_{IR}$, using the fact that, to the order we work in conformal perturbation theory, $\tilde R_{IR}$ is given by
\eqn\RIR{
\tilde R_{IR}=\tilde R_{UV}+{2\over3}U~.
}
In particular, using the fact that the scaling dimension, $D$, of an $\CN=1$ chiral primary is given by $D={3\over2}\tilde R$, we see that the change in dimension of some $\CN=1$ primary, $\CO$, is
\eqn\changedim{
\delta D(\CO)=D_{IR}(\CO)-D_{UV}(\CO)={1\over256p^2+9c(1+2p)^2}\Big(9c(1+2p)\cdot J_{\Phi}(\CO)+64p\cdot J_{(A_3, D_{2p})}(\CO)\Big)~.
}

In order to understand how the shifts in dimension \changedim\ affect our operators, it is clearly useful to be able to say something about $c$. While we do not know how to precisely compute $c$ in this example, we can find a sufficiently strong parametric lower bound on it. Indeed, we first observe from \scalingdims\ that the $(A_3, D_{2p})$ theory has an $\CO(p)$-dimensional Coulomb branch. As a result, if we turn on some generic Coulomb branch vevs, we should find $\CO(p)$ free $U(1)$ multiplets in the IR. This long distance theory has an $a$ anomaly of order $\CO(p)$. From the $a$-theorem \refs{\KomargodskiVJ,\KomargodskiXV} we therefore conclude that the $a$ anomaly of the $(A_3, D_{2p})$ theory is at least $\CO(p)$. Using the Hofman-Maldacena bounds \HofmanAR, we also see that
\eqn\cbound{
a,c\gtrsim\CO(p)~.
}
This analysis explains the two equivalent conditions given above in item {\bf(c)}.\foot{Note that \cbound\ does {\it not} imply $a\sim c$.}

Let us first examine the operators of type {\bf(i)}. These operators have $J_{\Phi}\ge0$ and $J_{(A_3, D_{2p})}\ge0$ (with at least one of these charges non-zero). Therefore, from \changedim\ we see that the dimensions of such operators increase, and we should be careful to check that slightly relevant UV operators (besides $\Phi\CO_{0,p-1}$ and $\Phi\CO_0$) don't become exactly marginal in the IR. To see this situation does not occur, first consider operators of the type $\CO_{k,\ell}$ with dimension close to three. Using \cbound\ and \changedim, we see that $\delta D\sim{1\over cp}\lesssim\CO(p^{-2})$. However, from \scalingdims\ we see that the barely relevant operators of this type have dimension that is smaller than three by $\CO(p^{-1})$. Therefore, the shift in the dimension is not enough to make the operator marginal in the IR (furthermore, even if such a marginal operator did exist in the IR, it would turn out to be marginally irrelevant since it breaks $U(1)_F$ with positive charge, and, as we will see below, there is no negative $U(1)_F$-charged marginal operator in the IR). Next consider operators of the type $\Phi\CO_{k,\ell}$ (with $k\ne0$ or $\ell\ne p-1$). From \changedim, it is clear that such operators are not marginal in the IR (they deviate from marginality by at least $\CO(p^{-1})$). A similar analysis applies to $\Phi^3$ and $\Phi^2\CO_{k,\ell}$.

Next, consider operators of type {\bf(ii)} and {\bf(iii)}. These primaries have non-zero $U(1)_R$ charge and non-zero $SU(2)_R$ spin. As a result, their $\CN=1$ chiral components should parameterize mixed branches of the $\CN=2$ moduli space.\foot{This logic breaks down when we have a primary that can be written as $Q^{2\alpha}\CO'Q^2_{\alpha}\hat\CO$, $\CO'(Q^2)^2\hat\CO$, or $\CO Q^{2\alpha}\CO'_{\alpha}$, for chiral operators $\CO'$, $\hat\CO$, and $\CO'_{\alpha}$. Note that these operators can also involve $\Phi$, but the $U(1)$ field strength, $W^2_{U(1)}$, which also appears in this sector is an uninteresting decoupled operator. Fortunately, all the operators mentioned in this footnote (with the exception of $W^2_{U(1)}$) are necessarily irrelevant in the UV and also irrelevant in the IR since $\delta D>0$ in \changedim. Note also that the results of \BuicanQLA\ imply that there are no $\CO'_{\alpha}$ in the $(A_3, D_{2p})$ theory.} However, mixed branches behave locally as products of Higgs branches and Coulomb branches since the vector multiplet and hypermultiplet moduli cannot mix in the K\"ahler potential \refs{\ArgyresEH} (see also the recent discussion in \XiePUA). Therefore, we expect such $\CN=2$ primaries to be products of $U(1)_R$-charged $SU(2)_R$ singlets and $SU(2)_R$-charged $U(1)_R$ singlets.\foot{Note that this discussion does not hold for $\CN=4$ theories, because such theories do not have an invariant distinction between the vector multiplet and hypermultiplet moduli spaces.} It is then easy to see that the operators with $j_R={1\over2}$ are of the form $Q\tilde\CO$, where $Q$ is part of a free hypermulitplet, and the operators with $j_R=1$ are of the form $\tilde\Phi\tilde\CO$, where $\tilde\Phi$ is part of a free $U(1)$ vector multiplet (note that we can have $\tilde\Phi=\Phi$). It then follows from \changedim\ that such operators are never marginal in the IR.\foot{Also, for the case of the hypermultiplet or $\tilde\Phi\ne\Phi$, applying \Za\ for the $J$ symmetries that act on the free vector and the free hypermultiplet respectively shows that such operators are never exactly marginal in the IR.} Let us also note that \cbound\ (and, more generally, condition {\bf(c)}) guarantees that an operator of type {\bf(iii)} having the form $\Phi\tilde\CO$ (with $\CO$ a holomorphic moment map for $G$) is indeed irrelevant in the IR (otherwise, if $c\sim\CO(p^0)$, it could have happened that this operator would have remained marginal in the IR with $J_F<0$).

Next, consider the type {\bf(iv)} operators.\foot{Examples of such operators include the baryons in $SU(3)$ SQCD with $N_f=6$.} Any such operators become relevant in the IR since they have $J=-3$. Note that this reasoning does not depend on knowing the precise number of these operators.

Let us now consider the $\CN=1$ primaries that are $\CN=2$ descendants. To that end, we first study type {\bf(v)} operators, $(Q^2)^2\CO$. Let us suppose that $\CO$ is from the $(A_3, D_{2p})$ sector. If $(Q^2)^2\CO$ is relevant in the UV, then it follows from the above discussion that $\delta D<0$, and this operator is more relevant in the IR. On the other hand, if $(Q^2)^2\CO$ is irrelevant in the UV, then it follows that $\delta D>0$ and the operator is more irrelevant in the IR. The potentially troublesome case is when $(Q^2)^2\CO$ is marginal in the UV (this is the usual $\CN=2$-preserving marginal deformation, and $\CO$ is the dimension two prepotential deformation). In this case, $(Q^2)^2\CO$ has dimension three in the IR as well. However, we see from \scalingdims\ that the $(A_3, D_{2p})$ theory has no $\CO$ of dimension two (i.e., it is isolated as an $\CN=2$ theory\foot{In fact, it is isolated as an $\CN=1$ theory as well.}). In particular, we see that it is crucial that our theory satisfies property {\bf(a)} given in the introduction. Finally, let us suppose that $\CO=\Phi\hat\CO$, where $\hat\CO$ is from the interacting UV sector. Such an operator is necessarily irrelevant and again $\delta D>0$, so the operator is more irrelevant in the IR (note also that contribution from $\Phi (Q^2)^2\hat\CO\subset (Q^2)^2(\Phi\CO)$ is guaranteed to be irrelevant by \cbound\ and, more generally, property {\bf(c)}).

Next let us consider the type {\bf(vi)} operators, $Q^{2\alpha}\CO_{\alpha}$. From \BuicanQLA, we know such operators do not exist in the $(A_3, D_{2p})$ theory. However, we can still argue that such operators do not yield new exactly marginal deformations in the IR even without using \BuicanQLA. Indeed, let us suppose that there is an $\CO_{\alpha}$ in the $(A_3, D_{2p})$ theory. In this case, the descendant can be at most marginally irrelevant in the IR since it breaks the $U(1)_F$ symmetry (with $J_F>0$). Finally, let us suppose that $\CO_{\alpha}=\Phi\hat\CO_{\alpha}$, with $\CO_{\alpha}$ from the $(A_3, D_{2p})$ theory. Such an operator is not relevant in the UV and is irrelevant in the IR, since it has $\delta D>0$. A similar analysis shows that type {\bf(vii)} operators are irrelevant in the UV and the IR.

Finally, one may wonder if higher-order perturbative corrections change the above picture, since the exact IR $R$-symmetry differs from the $R$-symmetry in \RIR. However, we can compute the exact IR $R$-symmetry using $a$-maximization \IntriligatorJJ. We find that the exact IR $R$-symmetry differs from the $R$-symmetry discussed above by 
\eqn\exactdiff{
\delta R\sim-{32\over27cp^2}J_F~.
}
Such corrections are highly subleading relative to \changedim\ and are not sufficient to make any of the non-marginal IR chiral primaries marginal (happily, since $\Phi\CO_0$ and $\Phi\CO_{0,p-1}$ are invariant under $J_F$, these higher-order corrections do not affect the dimensions of the exactly marginal deformations that take us along $\CM_c$). We conclude our discussion of this example by noting that we do not find any violations of unitarity bounds in the IR (at least in the chiral sector).

We can also generalize the previous example and consider any $(A_{2^n-1}, D_{2p})$ SCFT with $n\ge2$, $2^n\ll p$, and a free chiral multiplet (the case $n=1$ does not yield a non-trivial compact IR conformal manifold). The analysis for general $n$ is more tedious but proceeds largely as it does for $n=2$. In the end, we find an IR ${\bf C}P^1$ conformal manifold given by
\eqn\AgenCM{\eqalign{
\Big\{\lambda^i\Big||\lambda^1_*|^2+|\lambda^2_*|^2&={2p+2^{n-1}-1\over2^{n-1}}\cdot{\tau_U\over4\pi^4}={9\cdot2^{n-2}\cdot c\cdot(4p+2^n-2)\over9c(4p+2^n-2)^2+16(8p+2^n-4)^2}\cr&\lesssim{2^{n-4}\over \pi^4p}\ll1\Big\}\Big/U(1)~,
}}
where the $U(1)$ action is generated by $U$. We see that as $n$ increases, so too does the radius of the ${\bf C}P^1$.

One other difference for $n>2$ is that the $(A_{2^n-1}, D_{2p})$ theory may not be isolated as an $\CN=1$ theory (it is still isolated as an $\CN=2$ theory). In other words, there may be exactly marginal deformations in the UV that break $\CN=2\to\CN=1$. The reason this may happen is that for $n>2$, there are $\CN=2$ chiral primaries of dimension three (e.g., for $n=3$, the operator $\CO_{1,p-2}$ has dimension three). As a result, if there are primaries with $R_{\CN=2}=0$, $j_R={3\over2}$ (i.e., type {\bf(iv)} primaries), and appropriate quantum numbers under the $\CN=2$ flavor symmetry of the theory, then they may form exactly marginal linear combinations with the $\CN=2$ chiral primaries of dimension three. However, this possibility does not affect our analysis. Indeed, we find that the putative exactly marginal deformations in the UV break up into irrelevant and relevant deformations in the IR, and so any points on the UV conformal manifold that are close to the $\CN=2$ point are mapped to new IR fixed points with lower value of $a$ than the fixed points on the IR ${\bf C}P^1$.

\newsec{Comments on More General Compact Conformal Manifolds}
We have seen that we can naturally construct a ${\bf C}P^1$ conformal manifold in conformal perturbation theory via an $\CN=2\to\CN=1$ breaking relevant deformation that couples a free chiral multiplet, $\Phi$, to an $(A_{2^n-1}, D_{2p})$ AD theory. Furthermore, it is clear from our analysis that one potential way to construct theories realizing a ${\bf C}P^{N-1}$ conformal manifold for $N\ge3$ is to look for an $\CN=2$ SCFT satisfying properties {\bf(a)}-{\bf(c)} described in the introduction and then couple a free chiral multiplet to the corresponding operators described in {\bf(b)} via an interaction of the type given in \deformation. Note that we could also try to find a single sector theory with $N$ $\CN=2$ chiral operators of dimension $3-\epsilon$. We might then try to deform the superpotential by adding these operators. However, the one-loop fixed points in these cases are harder to control since the couplings get pushed to $\sum_i|\lambda|^2\epsilon\sim \epsilon^2c$. Of course, if $\epsilon c\ll1$, then these fixed points are under perturbative control. While we have not been able to find theories with this kind of scaling, we also have not found a proof that they do not exist.

As for constructing weighted projective space conformal manifolds of the type discussed below \satisfy, it is clear we cannot proceed via deformations of the type in \deformation. One natural way in the $\CN=2$ setup to accommodate a $U(1)$ action with many different charges is to deform the prepotential by operators of dimension $2-\epsilon_i$. However, this deformation preserves $\CN=2$, and the one-loop beta function is so small (due to bounds arising from the conformal bootstrap) that the theory is pushed to non-perturbatively large couplings \BuicanICA. Therefore, we should either use more sophisticated tools to analyze this case,\foot{The results in \refs{\PapadodimasEU,\GerchkovitzGTA,\BaggioIOA,\BaggioSNA, \GomisWOA} might be useful for such an analysis.} or we should find a different way to realize a weighted projective space: perhaps using $\CN=1$ theories. It would also be very interesting to understand if it is possible to construct more general projective varieties.

In order to construct more general toric varieties, we would need to have multiple broken $U(1)$ symmetries. We might find such examples with arbitrarily many broken $U(1)$ symmetries by coupling together different $\CN=2$ sectors with a chiral $\Phi$ multiplet via interactions of the type in \deformation. Alternatively, we could try to construct examples using deformations from additional sectors of operators of $\CN=2$ theories (i.e., operators charged under $\CN=2$ flavor symmetries) or perhaps using more general $\CN=1$ theories.

Another interesting direction would be to explore Grassmanian conformal manifolds in four dimensions. In order to construct such spaces, we need some broken non-abelian symmetry (as is the case in the three dimensional models of \ChangSG). One option for constructing such conformal manifolds might be to use other sectors of operators of $\CN=2$ theories that are charged under non-abelian flavor symmetries or more general $\CN=1$ theories. We hope to return to these and other more exotic conformal manifolds in the near future. 

Finally, it is interesting to ask how our above discussion works holographically. While we have not worked out all the details and implications, one can, for example, check that for the SCFTs in \XieJC\ that are candidates for theories with good supergravity duals (i.e., where one can compute $a$ and $c$ and find $a\sim c\gg1$), one does not find examples with compact conformal manifolds. More precisely, these theories violate at least one of our requirements {\bf(a)-(c)} mentioned in the introduction.

\bigskip\bigskip
\centerline{\bf Acknowledgements}
\noindent
We would like to thank K.~Papadodimas and Y.~Tachikawa for interesting discussions and correspondence. We are particularly grateful to Z.~Komargodski for many fruitful discussions on the topics discussed in this note. This work was partly supported by the U.S. Department of Energy under grants DOE-SC0010008, DOE-ARRA-SC0003883 and DOE-DE-SC0007897.

\listrefs
\end